\documentclass[
  DIV16,
  a4paper,
  11pt,
]{scrartcl}

\usepackage[english]{babel}
\usepackage{graphicx}
\usepackage{amsmath}
\usepackage[utf8]{inputenc}
\usepackage[T1]{fontenc}
\usepackage{mathptmx}
\usepackage{color}
\usepackage{array}
\usepackage{subfig}
\usepackage{url}
\usepackage{cite}
\usepackage{sidecap}

\usepackage{authblk}

\newcommand{\xv}[1]{\boldsymbol{#1}}

\newcommand{\fn}[1]{\textsuperscript{#1}}

\newcommand{\mccbf}[1]{\multicolumn{1}{c}{\textbf{#1}}}
\newcommand{\mccbmf}[1]{\multicolumn{1}{c}{\boldmath\ensuremath{#1}}}

\newcommand{\xlb}{B_l}
\newcommand{\mbf}{\frac{\text{B}}{\text{FLUP}}}
\newcommand{\xdpdf}{D_\text{pdf}{}}
\newcommand{\xdidx}{D_\text{idx}{}}
\newcommand{\xdblk}{D_\text{block}{}}

\newcommand{\xbosnt}{B_l(\text{OS-NT}){}}
\newcommand{\xbosntr}{B_l(\text{OS-NT-R}){}}

\newcommand{\xbaa}{B_l(\text{AA}){}}
\newcommand{\xbaar}{B_l(\text{AA-R}){}}

\newcommand{\xaa}{AA}

\newcommand{\xaaa}{AAOptAvx}

\newcommand{\ilbdc}{ILBDC}

\newcommand{\bi}{\begin{itemize}}
\newcommand{\ei}{\end{itemize}}
\newcommand{\be}{\begin{equation}}
\newcommand{\ee}{\end{equation}}

\usepackage[colorlinks=true, linkcolor=blue, citecolor=blue, urlcolor=blue, bookmarksopen=true, bookmarksnumbered=true]{hyperref}

\title{Modeling and analyzing performance for highly
optimized propagation steps of the lattice Boltzmann method on sparse lattices}
 
\author[*]{M. Wittmann}
\author[*]{T. Zeiser}
\author[*]{G. Hager}
\author[**]{G. Wellein}
\affil[*]{Erlangen Regional Computing Center, University of Erlangen-Nuremberg, Germany}
\affil[**]{Department of Computer Science, University of Erlangen-Nuremberg, Germany}
\date{\today}

\begin{document}

\maketitle

\begin{abstract}
  Computational fluid dynamics (CFD) requires a vast amount of compute
  cycles on contemporary large-scale parallel computers. Hence,
  performance optimization is a pivotal activity in this field of
  computational science.  Not only does it reduce the time to solution,
  but it also allows to minimize the energy consumption.  In this work we
  study performance optimizations for an MPI-parallel lattice
  Boltzmann-based flow solver that uses a sparse lattice
  representation with indirect addressing.  First we describe how this
  indirect addressing can be minimized in order to increase the single-core
  and chip-level performance.  Second, the communication overhead
  is reduced via appropriate partitioning, but maintaining the
  single core performance improvements.  Both optimizations allow to
  run the solver at an operating point with minimal energy
  consumption.
\end{abstract}

\section{Introduction and related work}

Performance optimization for computational
fluid dynamics (CFD)  should start at the core and chip level.
In most CFD algorithms the relevant bottleneck is the main memory bandwidth.
A high-quality implementation therefore should exhaust
the available bandwidth as well as cause the least amount of memory
data traffic per work unit.

In case of lattice Boltzmann methods (LBM)
the decisive quantity is called \emph{loop balance}
($\xlb$) and is measured in bytes per
fluid lattice site update ($\mbf$)~\cite{wittmann-2012-cam}.
There are several 
propagation step variants for LBM that achieve lowest data traffic per
FLUP: the two-grid one-step
algorithm with non-temporal stores~\cite{wellein-2006}, Bailey et.\ al's
AA-pattern~\cite{bailey-2009}, and Geier's Eso-Twist~\cite{linxweiler-2011,
schoenherr-2011}.
The latter two both work with a single grid only and arrange the
processing order in a clever way to work around data dependencies.
The first two variants were successfully implemented in the fluid flow
solver framework \ilbdc{}~\cite{wittmann-2013-sc13}, which relies on a sparse
lattice structure of the simulation domain~\cite{zeiser-2009-ppl}.
In contrast to a full array approach, this introduces indirect data accesses, but
delivers, if done correctly, excellent performance not only for flow in
simple geometries but also in porous media like fixed-bed reactors or foams.

In this work we present a modified version of the two-grid one-step
algorithm with non-temporal stores (OS-NT) and the AA-pattern algorithm.
Both are augmented by a technique called RIA (reduced indirect addressing),
which can avoid the indirect access under certain conditions.
This optimization is based on the idea of run length encoding and enables a
reduction of the loop balance $\xlb$. It also allows for partial vectorization, which is
usually incompatible with indirect addressing unless the hardware supports
efficient gather operations. RIA and partial vectorization were already 
implemented in the \ilbdc{} code we employed for our earlier 
analysis~\cite{wittmann-2013-sc13}; here we describe
them in due detail. 

Beyond the chip level we focus on optimizations regarding large scale parallel
execution.
The partitioning becomes more and more important with rising communication
overhead. The goal here is to reduce the communication volume and possibly
the number of neighbors of all partitions without impacting the single-core 
performance.
In case of \ilbdc{} partitioning is performed by cutting the adjacency list in
chunks of equal size.
The quality of the resulting partitions depends on how the adjacency list was
set up, i.e., which enumeration function was chosen (see~\cite{wittmann-2011-caf} 
for details).
Locality-preserving enumeration schemes such as space filling curves (SFC) or
lexicographic sorting (LS) with small blocking factors result in compact
partitions but have the disadvantage of poor single core
performance, because they lead to short loops and non-vectorizable code.
Better single core performance can be achieved with large blocking factors
for LS, which in turn yields partitions of lower quality.
The ideal solution here depends on the simulation geometry and requires
experimenting with the parameters to find a good point where both requirements
are met.
We propose a two-stage method to tackle this problem:
In the first step space filling curves are used for the adjacency list setup, which 
will generate high quality partitions.
After the partitions have been  assigned to the solver processes, they are re-enumerated
by LS (without a blocking factor). 
This leads to a partitioning with low communication overhead and high
single core performance. No manual experimentation, auto-tuning, etc.\ is 
required, and it is independent
of the number of partitions.

The paper is structured as follows.
In Sect.~\ref{sec:lbm} we give a brief introduction to the lattice
Boltzmann methods.
The compute cluster used for performance measurements is described in
Sect.~\ref{sec:tb}.
The principles of OS-NT and AA-pattern and their optimizations are discussed in
Sect.~\ref{sec:prop}.
These optimizations are evaluated with \ilbdc{} on
a current Intel Haswell processor in Sect.~\ref{sec:perf}.
In Sect.~\ref{sec:ls} large-scale performance is evaluated.
Finally we summarize the results in Sect.~\ref{sec:conclusion}.

\section{Lattice Boltzmann Methods}
\label{sec:lbm}

\begin{SCfigure}[1.3][tb]
\subfloat[Collision]{\label{fig:lbe:coll}\includegraphics[width=0.20\textwidth, clip=true]{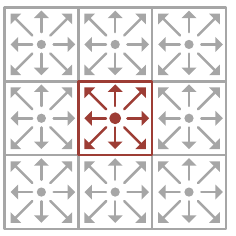}} \,
\subfloat[Propagation]{\label{fig:lbe:prop}\includegraphics[width=0.20\textwidth, clip=true]{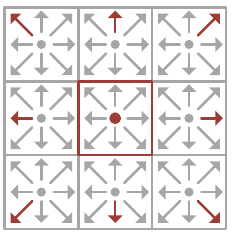}}
\caption{Collision~\protect\subref{fig:lbe:coll} of the PDFs of the centering 
node and the following propagation~\protect\subref{fig:lbe:prop} of the 
post-collision PDFs for a D$2$Q$9$ model.\medskip}
\label{fig:lbe}
\end{SCfigure}

Lattice Boltzmann Methods are derived from the lattice Boltzmann equation (LBE) and 
result from a discretization of the velocity space and a numerical discretization
of the spatial and time derivatives of the Boltzmann equation~\cite{succi-2001,wolf-gladrow}.
The discretization model is denoted D$d$Q$q$, where $d$ represents the spatial
dimension and $q$ is the velocity discretization.
Typical implementations are D$2$Q$9$, D$3$Q$15$, or D$3$Q$19$.
Each lattice node comprises $q$ particle distribution functions (PDF) $f_i$ with 
$i = 0, \dotsc, q -1$, which are the central element for computation.
The LBE is written as
\begin{equation}
  f_i(\xv{x} + \xv{c_i} \Delta t, t + \Delta t) = 
    f_i(\xv{x}, t) + \Omega(f_i(\xv{x}, t), f_i(\xv{x}, t)), \qquad 
    i = 0,\dotsc,q-1, 
  \label{eq:lbe}
\end{equation}
with the PDFs~$f_i$, the position vector $\xv{x}$, 
the velocity vector $\xv{c_i}$, and the collision operator $\Omega$.
The right-hand side of~\eqref{eq:lbe} describes the collision of PDFs of the current time step,
whereas on the left-hand side the propagation takes place, transferring the newly computed
post-collision PDFs to the neighbor nodes.
This is visualized in Fig.~\ref{fig:lbe} for a D$2$Q$9$ discretization.
In Fig.~\ref{fig:lbe:coll} the central node is about to be updated.
The read and collided PDFs are then propagated as depicted in 
Fig.~\ref{fig:lbe:prop}.
In this work the two-relaxation time (TRT) collision operator from 
\textsc{Ginzburg} et al.~\cite{ginzburg-2008} is used.

\section{Test Bed}
\label{sec:tb}

\begin{table}[tb]
\small
\centering
  \begin{minipage}{\linewidth}\centering
  \subfloat[]{
    \begin{tabular}{ll|rrrr}
      \hline
      \textbf{Processor} \\ 
      \hline
      Type            &&& Intel Xeon\\
                      &&& E5-2697 v3\\
      Base freq.      &[GHz]&& 2.6 \\
      Phys./SMT cores &&& 14/28 \\
      ISA ext.        &&& AVX, FMA3 \\
      \textit{Cache} &&& \\
      L1 (per core) &[KiB]&& 32, 8-ways \\
      L2 (per core) &[KiB]&& 256, 8-ways \\
      L3 (shared)\fn{a}  &[MiB]&& 2 $\times$ 17.5, 20-ways \\
      \textit{TLB} &&& \\
        L1 4 KiB pages  &&& 64 4-ways \\
        L1 2/4 MiB pages &&& 32 4-ways \\
        L2\fn{b} &&& 1024 8-ways\\
      \hline
    \end{tabular}
    \label{tab:tb:proc}
    } \,%
    \subfloat[]{
    \begin{tabular}{ll|rrrr}
      \hline
      \textbf{Compute node} \\
      \hline
        Sockets       &&& 2 \\
        Memory        &[GiB]&& $4 \times 16$ \\
        Cluster-on-Die &&& enabled\\
        NUMA LDs      &&& 4 (2 per CPU)\\
        NUMA policy  &&& default\\
        Huge Pages   &[MiB]&& 2 \\
      \hline
      \vspace{2.3cm} \\
    \end{tabular}
    \label{tab:tb:sys}
    }
    \caption{Specification of the processor~\protect\subref{tab:tb:proc} and
compute node~\protect\subref{tab:tb:sys} used for performance evaluation.}
    \label{tab:tb}
  \end{minipage}\vspace{1em}
  \begin{minipage}[t]{\linewidth}
    \footnoterule\footnotesize
    \textsuperscript{a}
       In Cluster-on-Die mode two times seven cores share $17.5$ MiB of L3 cache
       each.\par
    \textsuperscript{b}
       The L2 TLB is shared by $4$\,KiB and $2$/$4$\,MiB pages.\par
  \end{minipage}\hfill
\end{table}
The SuperMUC Phase~$2$
cluster\footnote{\url{http://www.lrz.de/services/compute/supermuc/}}, located at
the Leibniz Computing Center (LRZ) in Garching, Germany, was used as the
benchmarking platform.
This cluster is structured as six ``islands'' with $512$~compute nodes each.
The nodes are connected via InfiniBand FDR10, with a 
$4$:$1$ over-subscribed fat tree among the islands. 
One node comprises two Intel Xeon E5-2697 v3 processors with 14 cores
each.
The CPUs are based on the Haswell microarchitecture and the cluster-on-die
feature was enabled.
In this mode the processor exhibits two ccNUMA locality domains (LD) instead of
one, with seven cores per LD.
Further details are listed in Tab.~\ref{tab:tb} or can be found in
\cite{intel-orm-2015}.
We used the Intel C/Fortran compiler $15.02$ (with architecture-specific 
optimization parameters \texttt{-xCORE-AVX2} and \texttt{-fma})
and IBM MPI POE $1.4$.

\section{Optimized Propagation Step Implementations}
\label{sec:prop}

The performance of highly optimized LBM kernels is limited by the memory
bandwidth on current x86-based architectures.
If the discretization model and the collision operator are not
altered, i.e., if the number of FLOPs stays constant,
performance can thus only be increased by reducing the loop balance~$\xlb$ of a
loop kernel.

In~\cite{wittmann-2012-cam} we have shown that the two-grid one-step
algorithm with non-temporal stores~\cite{wellein-2006} (OS-NT), Bailey et.\ al's
AA-pattern~\cite{bailey-2009}, and Geier's Eso-Twist~\cite{linxweiler-2011,
schoenherr-2011} have the lowest loop balance.
Here we concentrate on the first two, which are compatible with the 
code structure of \ilbdc.
In the following we assume a D$3$Q$19$
discretization model and use double-precision arithmetic.

\subsection*{Loop Balance}

\begin{figure}[tb]
\centering
    \subfloat[Indirect Addressing]{%
      \label{fig:ia}%
      \includegraphics[width=0.48\textwidth, clip=true]{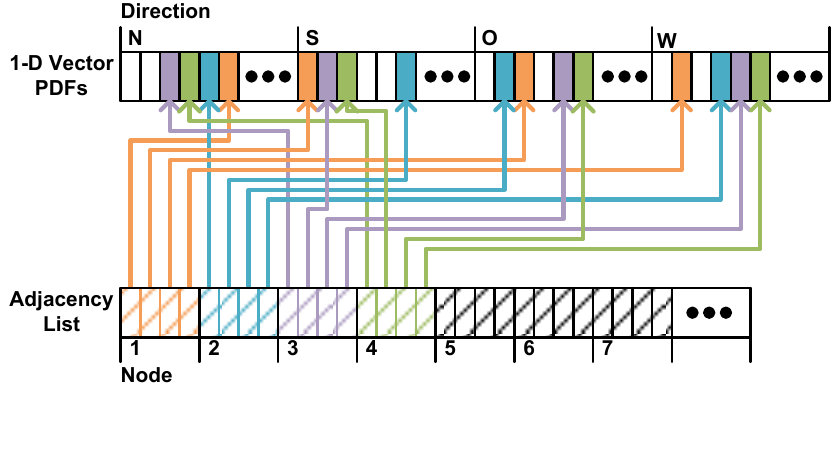}%
    }%
    \hfill%
    \subfloat[Reduced Indirect Addressing]{%
      \label{fig:ria}%
      \includegraphics[width=0.48\textwidth, clip=true]{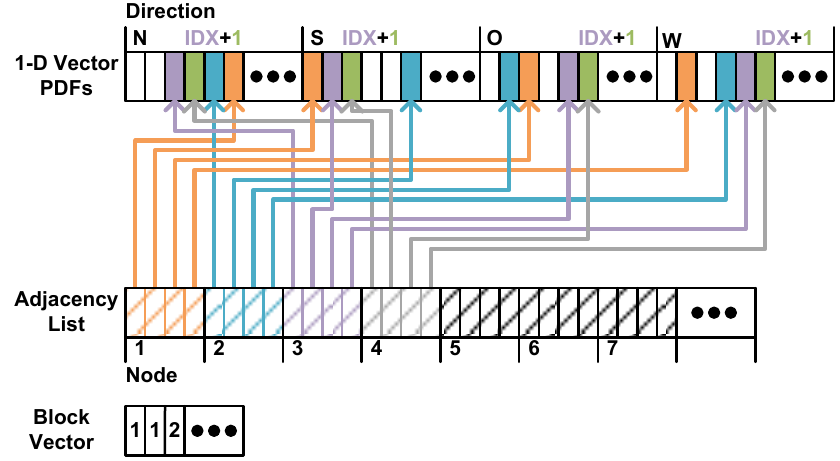}%
    }%
    \caption{%
      \protect\subref{fig:ia} Indirect addressing the PDF neighbors inside the 
      1-D vector via the adjacency list. 
      \protect\subref{fig:ria} Indirect addressing of certain PDF neighbors can 
      be avoided if the access pattern of the current node is the same as the
      previous one.
      This is the case for node $4$.
    }
    \label{fig:ia-all}
\end{figure}

Independently of the propagation step during each update of a fluid node $19$
PDFs must be loaded and stored, causing a data volume of 
\be
  \xdpdf = 2 \times 19 \times 8~\text{B} = 304~\text{B}. 
\ee
When a PDF must be accessed indirectly, its index in the adjacency list 
must be obtained first (see Fig.~\ref{fig:ia} for a visualization
of D$2$Q$5$). This causes an additional data volume of
\be
  \xdidx = 18 \times 4~\text{B} = 72~\text{B},
\ee
assuming a $4$-byte index.  Note that only $18$ instead of $19$
indirect accesses are needed, as the center PDF can always be directly
addressed.

The OS-NT algorithm with loop splitting employs the ``pull'' scheme:
First, PDFs from neighboring nodes are read via indirect addressing,
collided, and the post-collision values are stored back to the local 
node via a direct access:
\be
  \xbosnt = \frac{\xdpdf + \xdidx}{1~\text{FLUP}} = 376~\mbf{}.
\ee
Loop splitting, which is required to reduce the number of concurrent
non-temporal store streams, has no impact on the loop balance, as only data
traffic between cores and memory is
considered~\cite{wellein-2006,zeiser-2009-ppl}
The AA pattern requires direct access in the even time step ($\xdpdf$) and
indirect access in the odd time step ($\xdpdf + \xdidx$).
The average over both time steps is then
\be
  \xbaa = \frac{\xdpdf + \xdidx / 2}{1~\text{FLUP}} = 340~\mbf{}.
\ee

\subsection{Reduced Indirect Addressing}
\label{sec:prop:ria}

Reduced indirect addressing (RIA) is based on the idea that an indirect access
is not necessary for a fluid node if the previous fluid node exhibits the
same access pattern, but shifted by one.
Then, the pointers to the PDFs determined at the previous fluid node need only
be incremented to get the current node's PDFs.
As a consequence, the loop balance with RIA depends on how often the indirect access can be
avoided and is thereby dependent on the structure of the simulation geometry and
the chosen enumeration function to set up the adjacency list.
The information about how many consecutive fluid nodes share the same access pattern
(shifted by one) is stored as a run length coded (block) vector, as shown in
Fig.~\ref{fig:ria} for a D$2$Q$5$ discretization.
Each block vector access requires $\xdblk = 4$~B.
One can calculate the minimum and maximum loop balance for both propagation
patterns:
The lower and upper loop balance bounds for the OS-NT algorithm with RIA (OS-NT-R) are
\be
  \xbosntr{} = \left[
    \frac{\xdpdf}{1~\text{FLUP}}, 
    \frac{\xdpdf + \xdidx + \xdblk}{1~\text{FLUP}} \right] = 
  [ 304 , 380 ]~\mbf{}, 
\ee
while for the AA pattern with RIA (AA-R) we get
\be
  \xbaar{} = \left[\frac{\xdpdf}{1~\text{FLUP}}, 
    \frac{\xdpdf + (\xdidx + \xdblk)/2}{1~\text{FLUP}} \right] = 
  [ 304 , 342 ]~\mbf{}.
\ee

\subsection{Partial Vectorization}
\label{sec:prop:pv}

During the odd time step of \xaa{} the reading, writing, and collision of the PDFs
is still performed in scalar mode.
Via run length coding it is possible to \textit{partially} SIMD-vectorize
the  fluid node updates if at least $V$ nodes share the same access 
pattern, where $V$ is the vector width (in case of AVX, $V=4$).
As the optimized LBM kernels are memory bound and this optimization does not
alter the loop balance, no direct performance impact is expected when the
kernel is in the memory bound regime, i.e., when sufficient cores are
used to saturate the memory bandwidth.
Hence, the loop balance for AA-RP, i.e., AA with RIA and partial vectorization,
is still $\xbaar$.
However, we will show later that this optimization saturates the bandwidth with 
fewer cores, potentially saving energy.

\subsection{Architectural Optimizations}
\label{sec:prop:arch}

Due to the high number of concurrent memory streams ($38$ with OS-NT and $19$
with \xaa{}), D$3$Q$19$ bears the danger of cache thrashing.
In order to avoid this problem we introduce a padding into the data vector,
ensuring that different directions are always mapped on different sets of the
L1 data cache.
For OS-NT with $38$ concurrent memory streams the L1 TLB with $32$ entries for 
$2$~MiB pages is not large enough, leading to cache thrashing.
In contrast to previous Intel micro-architectures, a Haswell core provides an L2
TLB also for $2$~MiB pages, which mitigates the effect.
Although \xaa{} requires only $19$ concurrent entries in the L1 TLB for $2$~MiB
pages, thrashing can still occur if more than four page table entries are 
mapped to the same set.
Whether this happens depends on the number of ghost and fluid nodes.
To generally avoid this problem we use an additional padding against TLB
thrashing.
Details about the impact of thrashing on performance can be found 
in~\cite{wittmann-2016-phd}.

\section{Performance Results}
\label{sec:perf}

In this section we analyze the performance characteristics within one ccNUMA LD,
i.e., up to seven cores.

\subsection{Benchmark Geometries}
\label{sec:perf:geo}

Two different simulation geometries are employed for benchmarking.
The first is an empty channel with a quadratic cross section and dimensions
of $500\times100\times100$ nodes containing around $4.8\times10^6$ fluid nodes.
Due to the partial vectorization of the odd time step in AA, $98$\,\%
of the fluid nodes can be updated with SIMD instructions.
The second geometry originates from a real-world application and represents a
fixed-bed reactor of the same dimensions, but with only $2.1\times10^6$ fluid 
nodes.
The reason for the low partial vectorizablility of only $57$\,\% is the high
porosity of this setup.
The loop balance for both propagation step implementations 
is listed in Table~\ref{tab:lbs}.

\begin{table}[tb]
\small
\centering
  \begin{tabular}{ll|cccc}
    \hline
    \textbf{Geometry} &&  \mccbmf{\xbosnt} & \mccbmf{\xbosntr} & 
                          \mccbmf{\xbaa}    & \mccbmf{\xbaar} \\
    \hline
    Channel           &&   376 & 306 & 340 & 305 \\
    Fixed-bed reactor &&   376 & 333 & 340 & 319 \\
    \hline
  \end{tabular}
  \caption{Loop balance (in B/FLUP) of the benchmark geometries for the OS-NT and
           AA-Pattern propagation step implementations with and without RIA.}
  \label{tab:lbs}
\end{table}

\subsection{Roofline Performance Model}
\label{sec:perf:rm}

We use the Roofline performance model~\cite{williams-2009} to calculate upper
performance limits on the chip level, considering only the memory-bound case.
Typically the result of the  STREAM copy benchmark~\cite{mccalpin-1995} is used
as a bandwidth limit.
Here the variant with non-temporal stores achieved the highest bandwidth, shown
for one LD (seven cores) in Table~\ref{tab:bws}.
As shown in~\cite{habich-2012}, the prediction of the Roofline model for LBM
kernels can be improved if the streaming benchmark matches the access
pattern of the application.
Therefore we utilize two different micro-benchmarks for OS-NT and AA-pattern:
CNT-19A has the same characteristics as OS-NT.
$19$ arrays are copied chunk-wise into a small temporary array, from which data
is moved via nine loops, each copying two chunks utilizing non-temporal stores,
to the final array.
With U-19A the memory access pattern of the even time step of the
AA-pattern is mimicked by updating $19$ arrays concurrently.
The measured bandwidths of both benchmarks are given in Table~\ref{tab:bws}.
For reference we also list the bandwidths attainable by copying (with
non-temporal stores) and updating one array, named CNT-1A and U-1A,
respectively.
As already mentioned the saturated memory bandwidth on the Haswell system is
rather independent of the core frequency.
Consequently we see only a very small difference between
the bandwidths at $1.2$\,GHz and $2.6$\,GHz in Table~\ref{tab:bws}.
The final predictions of the Roofline model can be found in Table~\ref{tab:rm}.
They are visible as horizontal bars in Fig.~\ref{fig:p:p}.

\begin{table}[tb]
  \small
  \centering
  \begin{tabular}{l|rrrr}
    \hline
    Frequency   &   CNT-1A &    CNT-19A    & U-1A &  U-19A \\
    \hline
    1.2 GHz&  27.2   &     24.0 &  26.7 &  24.8       \\
    2.6 GHz&  27.2   &     24.0 &  26.7 &  25.1      \\ 
    \hline
  \end{tabular}
  \caption{Measured memory bandwidths (in GB/s) with different micro-benchmarks on seven
      cores of the benchmarked Haswell system, i.\,e.\ in one ccNUMA locality domain.
  }
  \label{tab:bws}
\end{table}

\begin{table}[tb]
  \small
  \centering
  \begin{tabular}{ll|rrrrrrrr}
    \hline
    &&& \multicolumn{3}{c}{\textbf{OS-NT}} && \multicolumn{3}{c}{\textbf{AA}} \\
    \cline{4-6} \cline{8-10}
    &&& w/o RIA && w/ RIA && w/o RIA && w/ RIA \\
    \hline
    \textit{Channel}\\
    \hline
    Loop Balance &[B/FLUP]&& 376 && 306 && 340 && 305 \\
    P @ $1.2$\,GHz   &[MFLUP/s] &&  63.8 && 78.4 && 72.9 && 81.3 \\
    P @ $2.6$\,GHz   &[MFLUP/s] &&  63.8 && 78.4 && 73.8 && 82.2 \\
    \hline
    \textit{Fixed-Bed Reactor}\\
    \hline
    Loop Balance &[B/FLUP]&& 376 && 333 && 340 && 319 \\
    P @ $1.2$\,GHz &[MFLUP/s] && 63.8&&  72.0 &&   72.9&&  77.8\\
    P @ $2.6$\,GHz &[MFLUP/s] && 63.8&&  72.0 &&   73.8&&  78.8\\
    \hline
  \end{tabular}
  \caption{Performance prediction of the Roofline model when propagation step
specific bandwidth micro-benchmarks are used.}
  \label{tab:rm}
\end{table}

\subsection{ECM Performance Model}
\label{sec:perf:ecm}

\begin{table}
  \small
  \centering
  \begin{tabular}{llllrrrrr}
    \hline
    \mccbf{port} && \mccbf{instructions} && \mccbf{ET} && \mccbf{OTB} && \mccbf{OTW} \\
    \hline
    0      && FMA, MUL && 146 && 144 &&  480 \\ 
    1      && FMA, ADD && 172 && 174 && 1080 \\
    2 \& 3 && LD       && 110 && 129 &&  460 \\
    4      && ST       &&  72 && 114 &&  360 \\
    \hline
  \end{tabular}
  \caption{Distribution of the in-core execution time in cycles on the core execution 
    ports for updating eight fluid nodes with AVX-vectorized AA-RP.
    Predictions are given for the even time step (ET) and the odd time step 
    in the best case (OTB) and worst case (OTW).
  }
  \label{tab:ecm}
\end{table}

The Execution-Cache-Memory (ECM) performance model~\cite{hager-2012-ecm} 
considers each layer of the memory hierarchy separately and
distinguishes between data transfers and arithmetic.
For brevity we only model the AVX-vectorized AA-RB here.
Within the ECM model cache line transfers play a major role.
Thus the modeling is done at a granularity of eight fluid node updates, as
(at least in the even time step) full cache lines are required.
The in-core execution time predictions were determined 
via IACA~\cite{intel-iaca} in throughput analysis mode,
which assumes that full overlap of instructions is possible
by out-of-order execution.
The even time step (ET) could be analyzed without modifications.
The kernel of the odd time step (OT) contains two conditional branches,
which are responsible for deciding about indirect access and SIMD
vectorization.
IACA only considers branches from loop iterations, but not branches
resulting from if conditions in the loop body.
Therefore two kernels representing the best and worst cases of OT described in
Sect.~\ref{sec:prop:ria} were built. 
The best case (OTB) contains only direct accesses with vectorized execution,
whereas in the worst case (OTW) only indirect access and scalar execution
occurs.
The results of the analysis are shown in Table~\ref{tab:ecm}.

For modeling the data paths the following bandwidths (in cycles per cache line,
cy/cl) between adjacent cache/memory levels are assumed: $1$~cy/cl between L1/L2,
$2$~cy/cl between L2/L3, $3.1$~cy/cl at $1.2$~GHz and $6.6$~cy/cl at $2.6$~GHz
between L3/memory. The latter two values are calculated from the
saturated memory bandwidth measurements (see above), while all others
are documented by Intel. 

\begin{figure}[tb]
  \centering
  \fbox{\includegraphics[width=0.95\textwidth, clip=true]{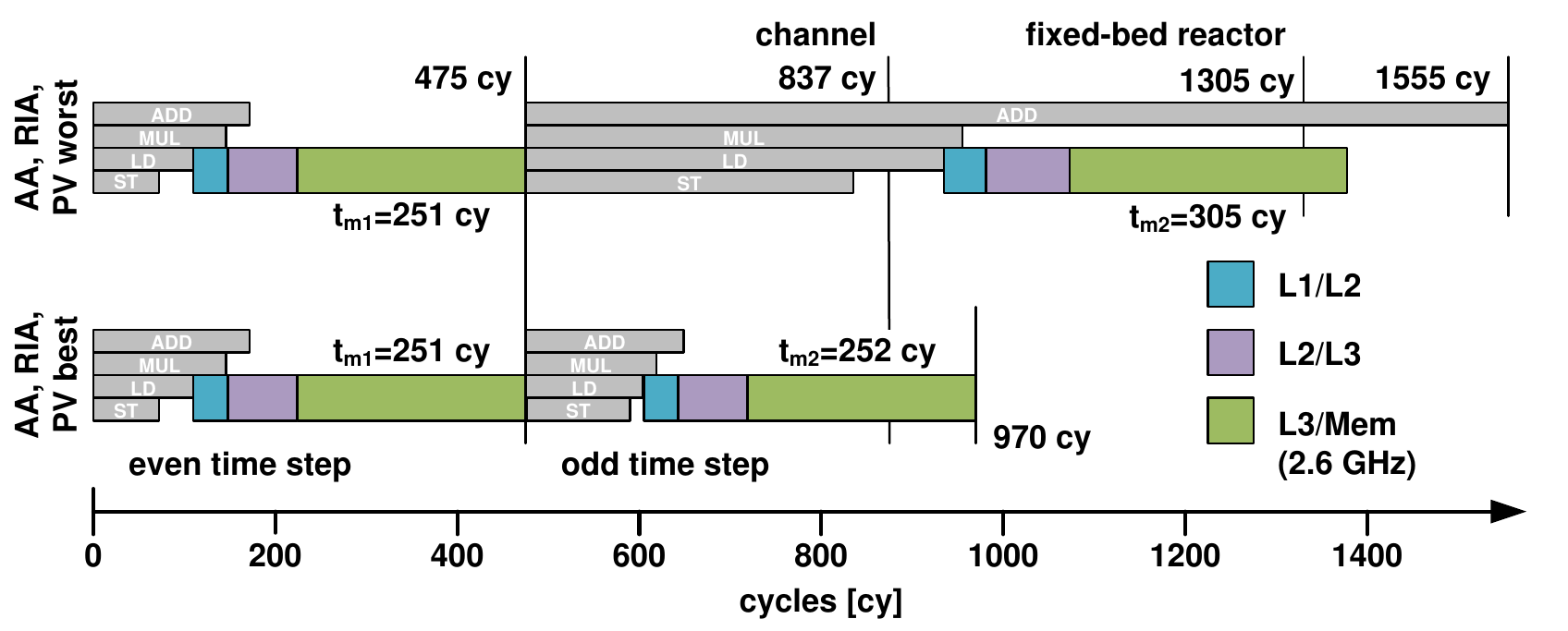}}
  \caption{Update of eight fluid nodes on one Haswell core at $2.6$~GHz modeled
with the ECM model for the even and odd time step of \xaa{} with RIA for best
and worst case.}
  \label{fig:perf:ecm}
\end{figure}

To update eight fluid nodes during the even time step,
$2\times 19$ cache lines must be transferred. This is also the case for OTB.
OTW requires additional traffic per fluid node: 
$8 \times 4$~B for the indirect access and $4$~B for the block vector.
For the work unit of eight fluid nodes this amounts to $4.5$ cache
lines. The execution diagram for the best and worst case is shown in
Fig.~\ref{fig:perf:ecm}.

ECM assumes linear performance scaling across cores until a bottleneck is reached.
As the Haswell CPU has a scalable L3 cache, the memory bandwidth is the
only potential bottleneck.
The resulting prediction are depicted as green lines in Fig.~\ref{fig:p:p}.

\subsection{Performance Results}
\label{sec:perf:results}

\begin{figure}[tb]
  \centering%
    \subfloat{%
      \label{fig:p:pa}%
      \includegraphics[width=0.45\textwidth, clip=true]{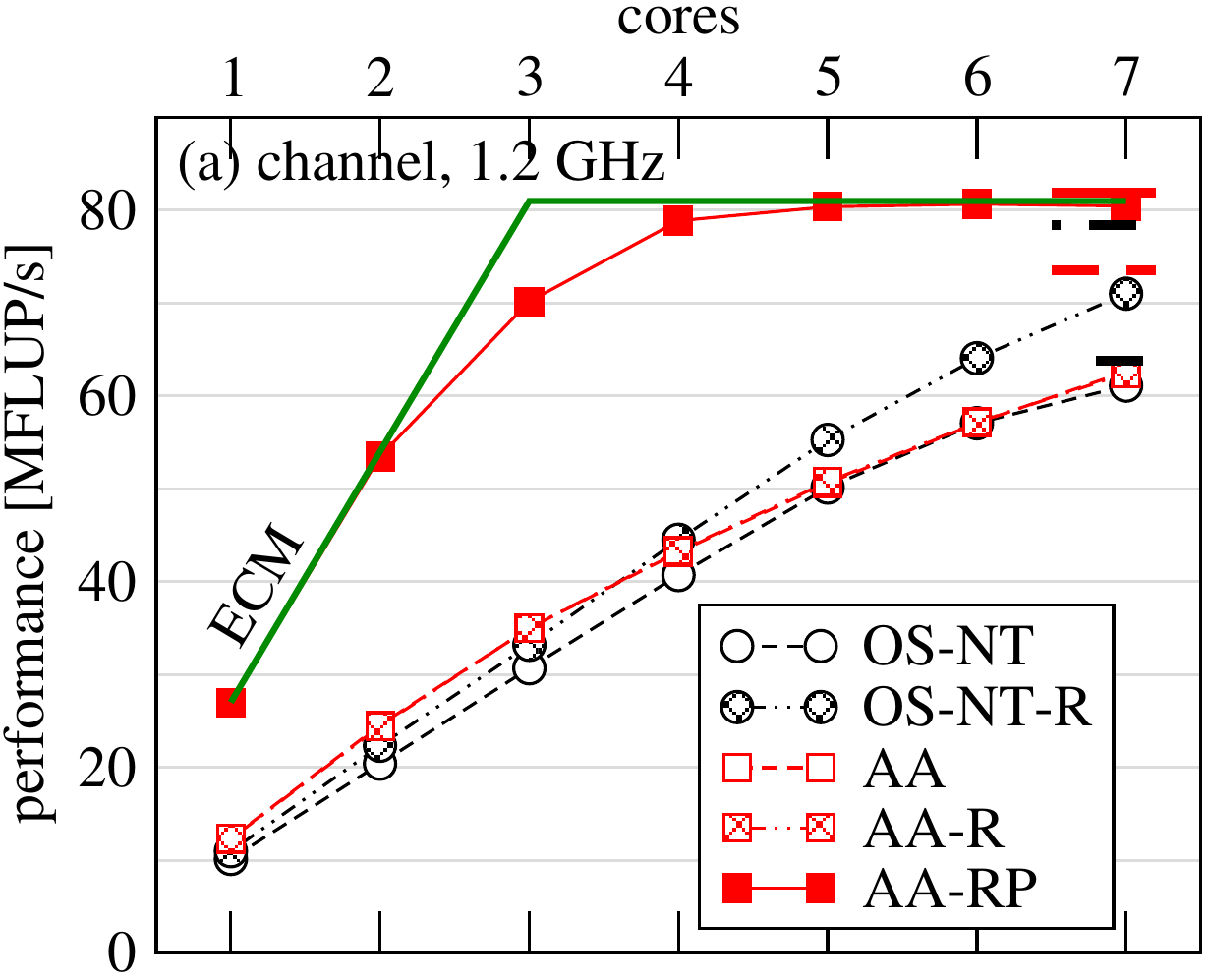}%
    }%
    \subfloat{%
      \label{fig:p:pb}%
      \includegraphics[width=0.45\textwidth, clip=true]{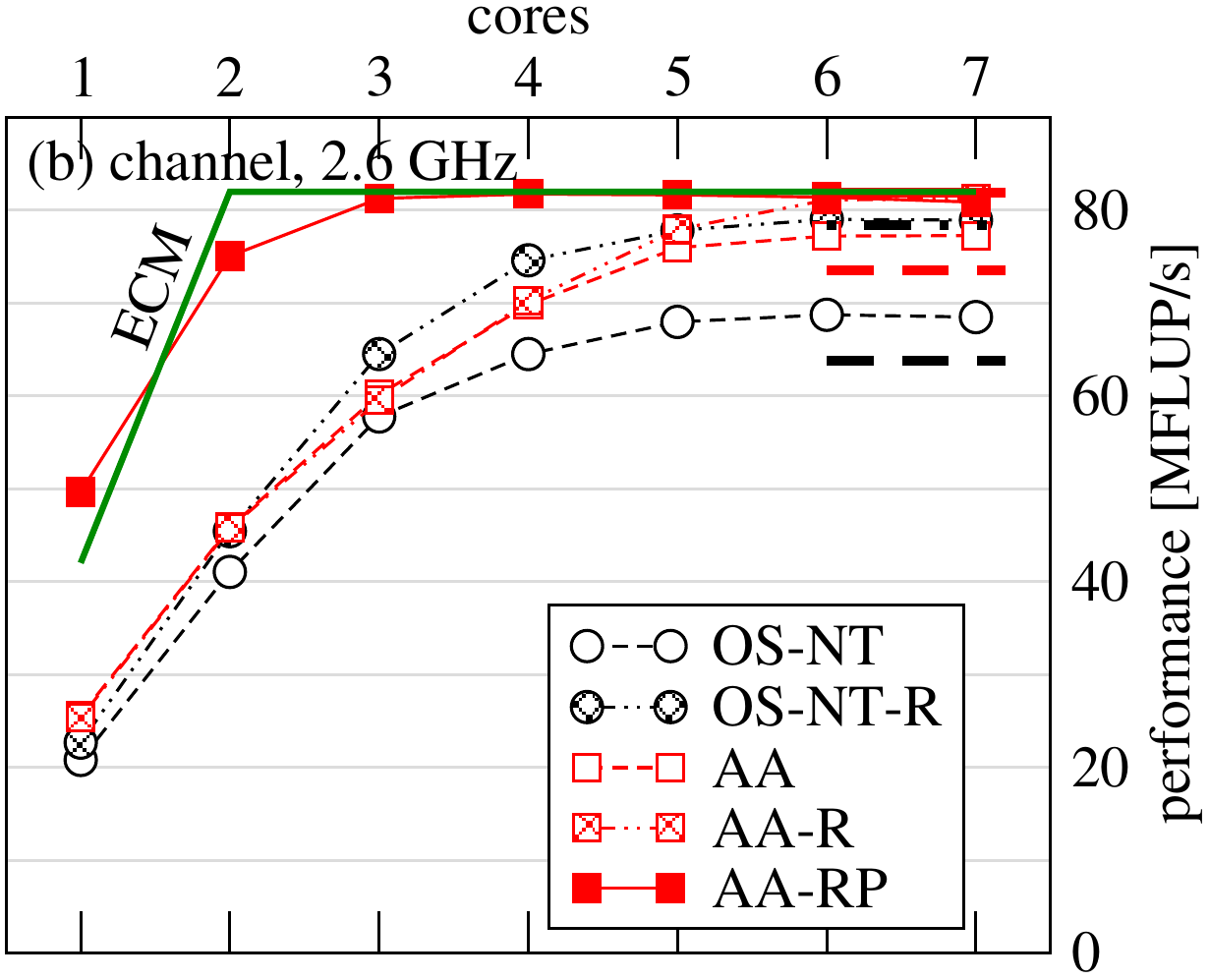}%
    }\\[-5mm]
    \subfloat{%
      \label{fig:p:pc}%
      \includegraphics[width=0.45\textwidth, clip=true]{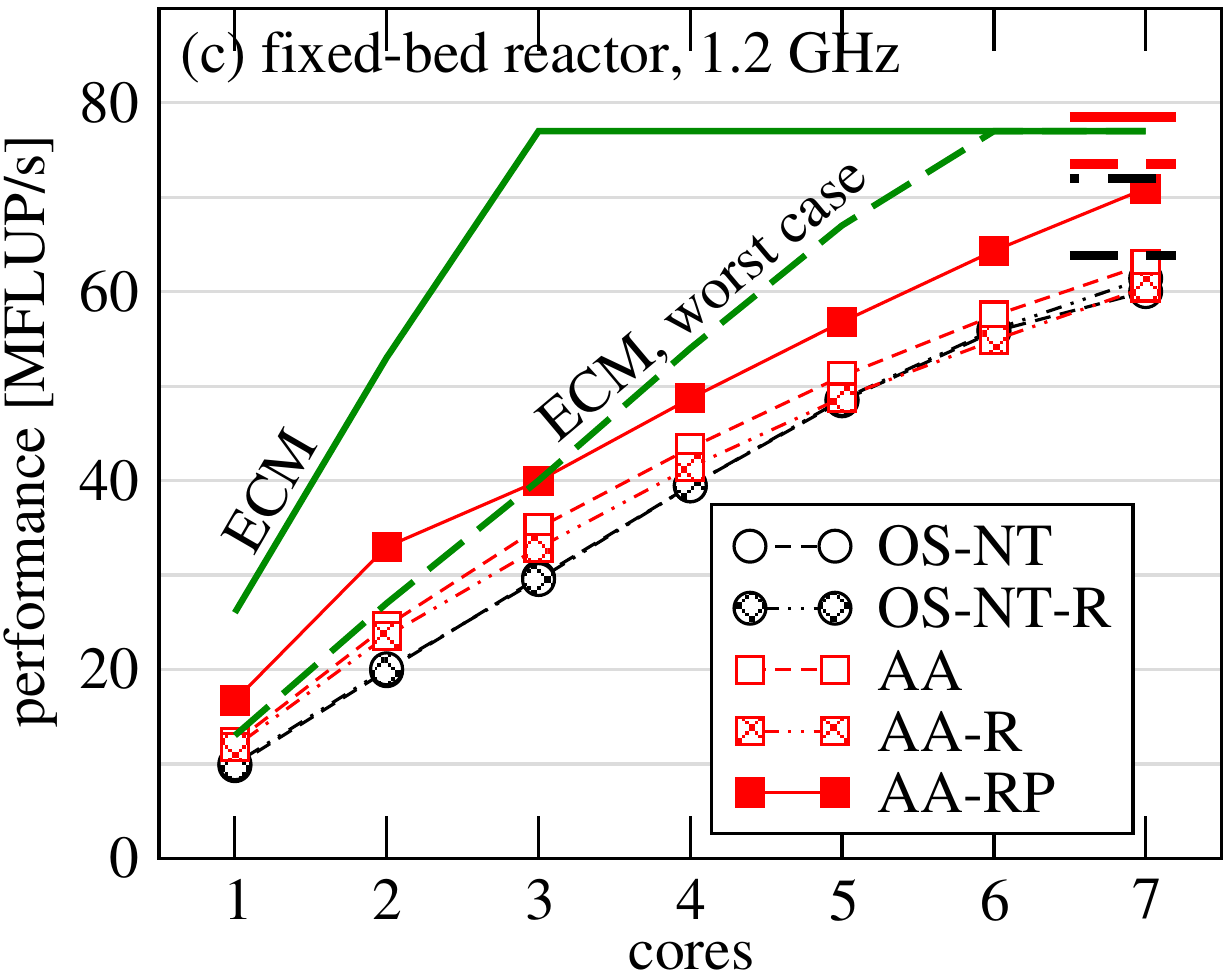}%
    }%
    \subfloat{%
      \label{fig:p:pd}%
      \includegraphics[width=0.45\textwidth, clip=true]{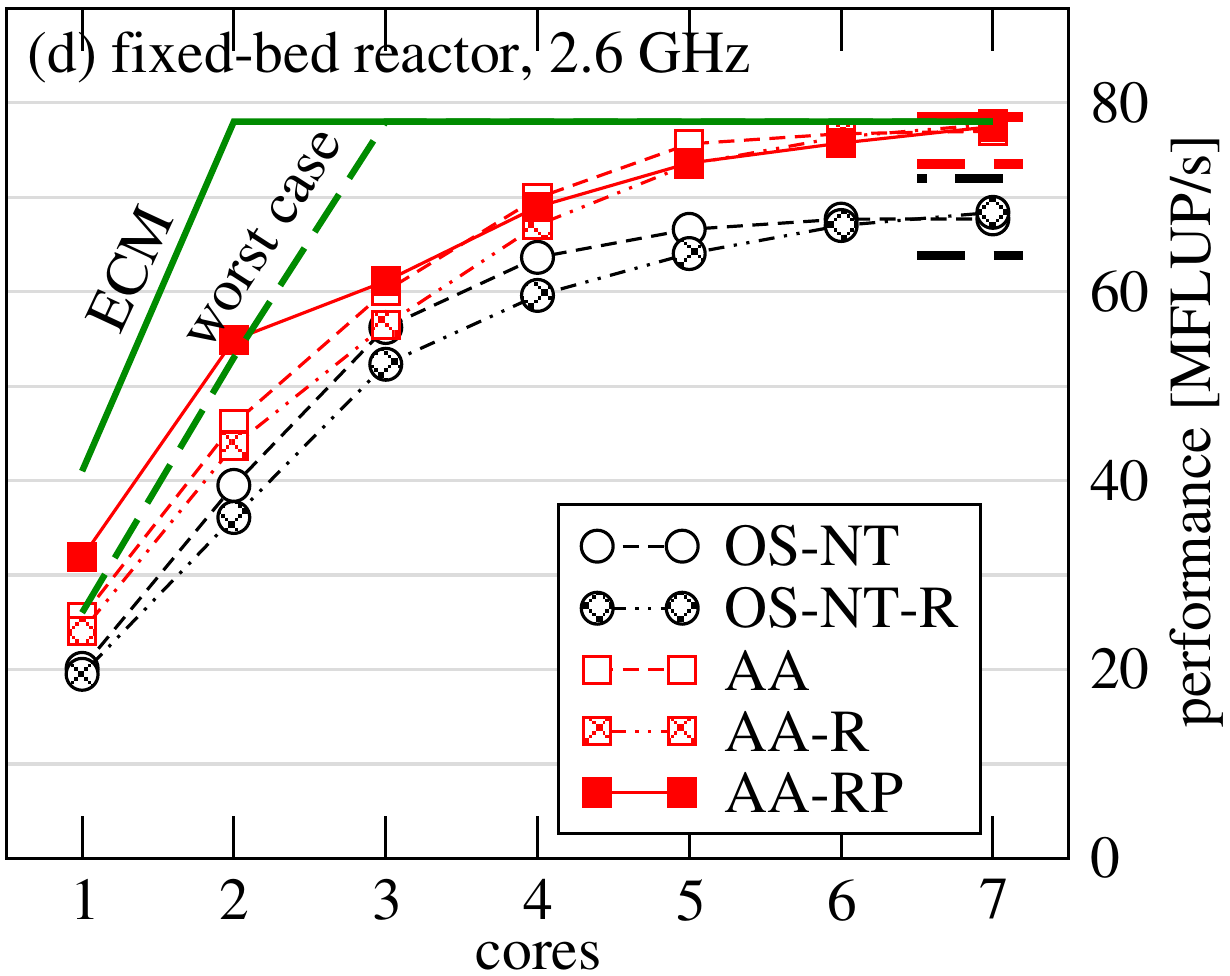}%
    }%
    \caption{Performance of OS-NT and AA on one Haswell NUMA LD without and with
RIA, respectively. Roofline limits (horizontal bars) for seven cores as well
as the prediction of the ECM model for the best (green continuous line) and
worst (green dashed line) case are shown as well.
    }
    \label{fig:p:p}
\end{figure}

The flow solver is purely MPI parallel. Each MPI process was pinned to 
a separate physical core, so the memory allocated by each process was ensured to
reside inside the core's LD. Performance results are shown
in Fig.~\ref{fig:p:p}.

The performance difference between the implementations at $1.2$\,GHz
and $2.6$\,GHz is only around $10$\,\%, independent of the simulation
geometry.
As with a lower frequency more cores are needed to saturate the memory
bandwidth, this also holds true for the benchmarked propagation steps.
Especially with the channel geometry and OS-NT(-R) as well as AA(-R),
performance saturation already occurs with fewer than seven cores at the 
higher frequency setting.
For AA-RP, where at both frequencies a saturation occurs at 
four cores already.
This is a consequence of the partial vectorization, which already boosts the single core
performance to over twice the level reached with AA and RIA alone.

The predictions of the Roofline model at $2.6$\,GHz are too pessimistic as OS-NT
and also AA exceed the predicted performance.
We think that the synthetic micro-benchmarks deliver a lower
bandwidth than the application code, as they have 
no precautions against cache/TLB thrashing.
According to the loop balance numbers in Table~\ref{tab:rm}, OS-NT-R (AA-R) should reach a
$20$\% ($10$\%) higher performance for the channel geometry compared to the
propagation step implementations without RIA, but only $15$\% ($5$\%)
improvement are actually achieved at $2.6$~GHz.
For the fixed-bed reactor the loop balances are only $10$\% ($6$\%) lower
with RIA than without for OS-NT (AA).
Here both OS-NT-R and AA-RP see no benefit at $2.6$~GHz.

The performance results for the propagation steps for the fixed-bed reactor, when
all seven cores are utilized, are at most only $10$\% lower than with the
channel geometry.
In this case AA-RP does not increase performance that much at low core
counts.
The reason is that for this geometry only $57$\% of the fluid cells (on average)
can be updated in a vectorized way during the odd time step.
Hence the impact on the single core performance is lower.

The ECM model for the best case assumption matches well the 
AA-RB result on the channel geometry, because of its high vectorizablility.
Here the model curves (green continuous lines) in Fig.~\ref{fig:p:p} (a) and (b)
nearly agree perfectly with the measured performance.
Because of the low vectorizability the fixed-bed reactor performance should be
close to the worst case assumption of ECM (green dashed lines).
This is only the case for up to two/three cores, shown in Fig.~\ref{fig:p:p} (c)
and (d).
With more cores the performance is lower than the prediction.
This behavior is clearly caused by effects that are not part of the model;
one candidate is misprediction of the conditional branches 
in the odd time step, caused by the small portions of fluid between
the obstacles. 

\subsection{Energy Efficiency}
\label{sec:perf:energy}

\begin{figure}[tb]
\centering%
\subfloat{%
\label{fig:p:e1}%
\includegraphics[width=0.45\textwidth, clip=true]{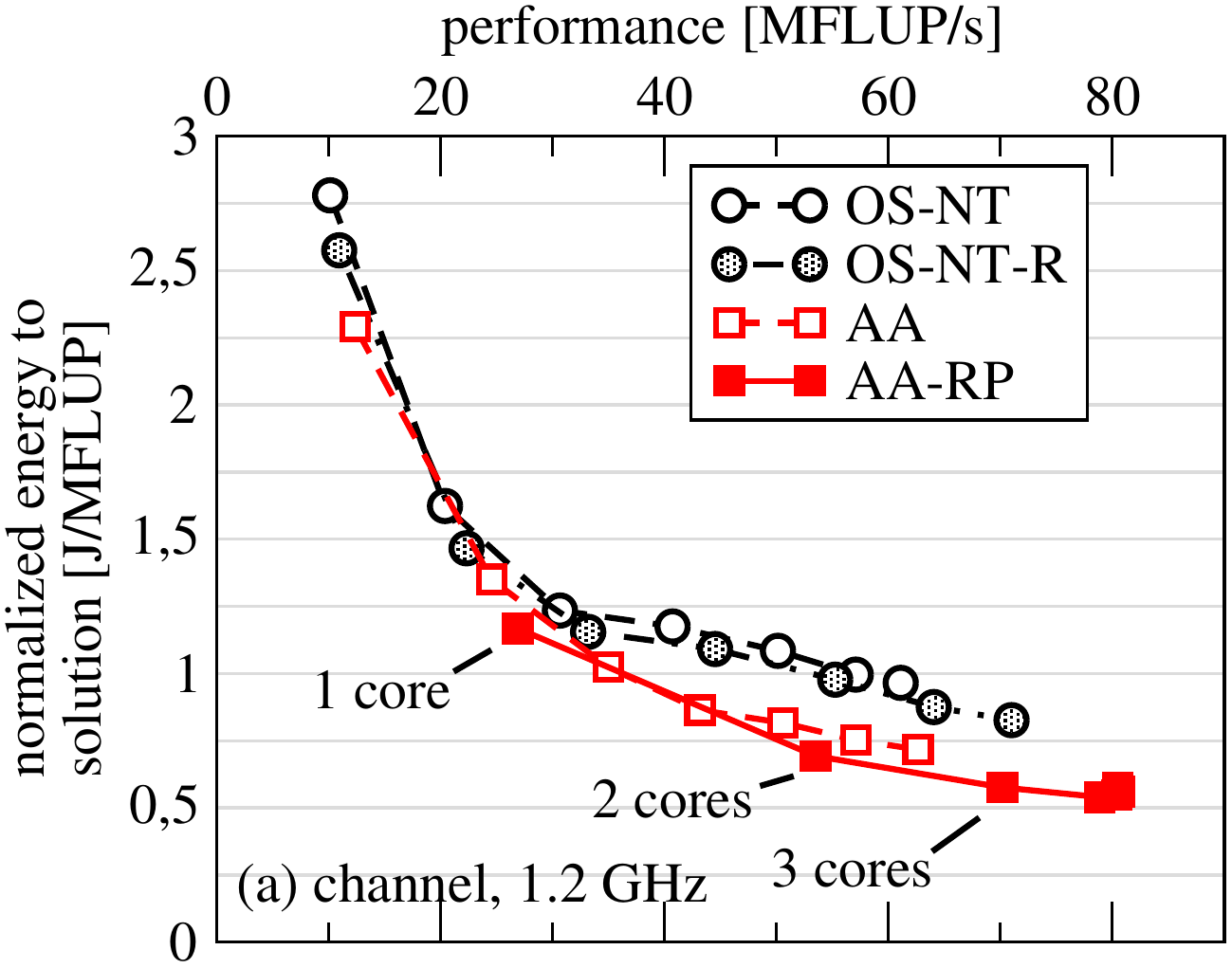}%
}
\subfloat{%
\label{fig:p:e2}%
\includegraphics[width=0.45\textwidth, clip=true]{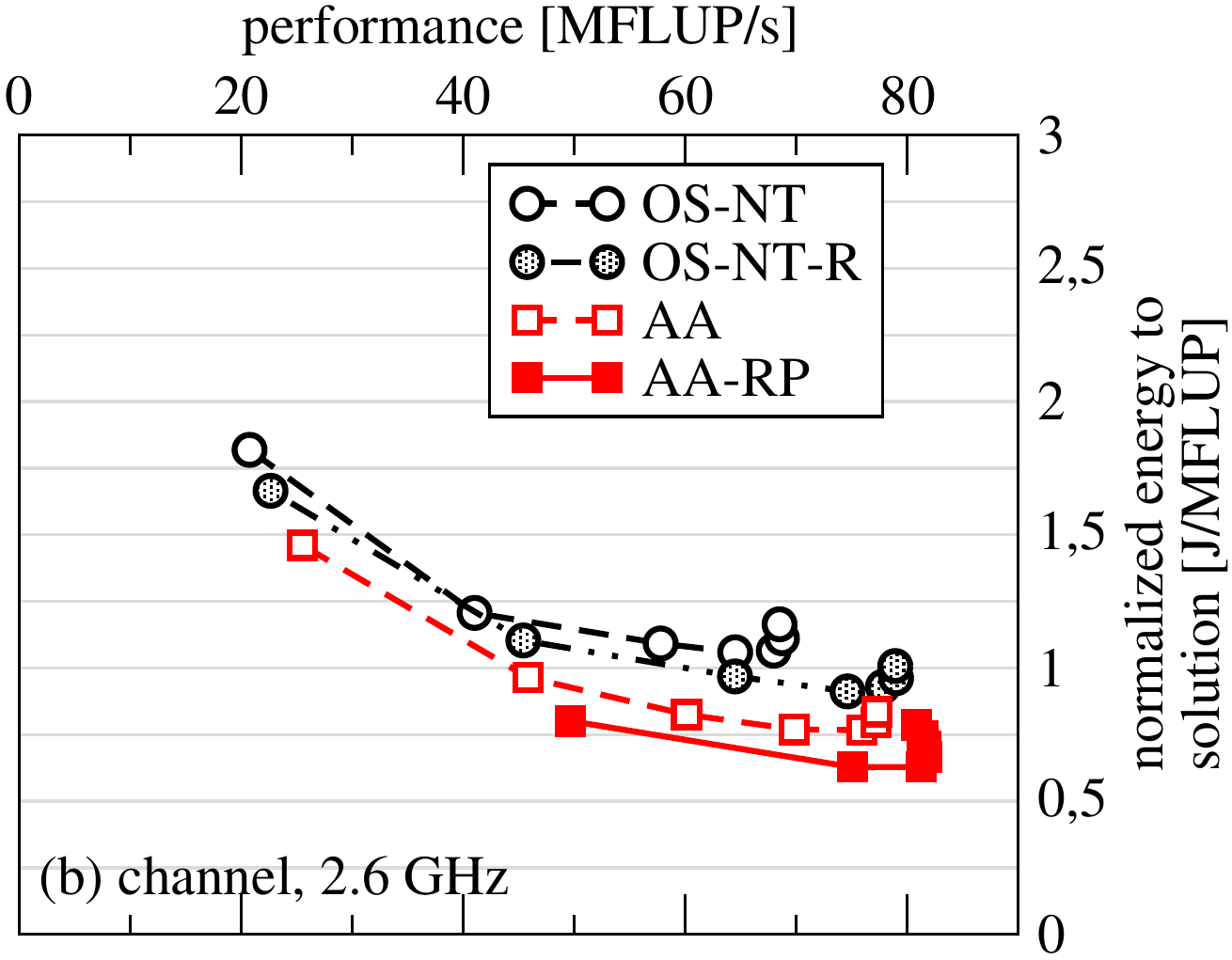}%
}\\[-4mm]  
\subfloat{%
\label{fig:p:e3}%
\includegraphics[width=0.45\textwidth, clip=true]{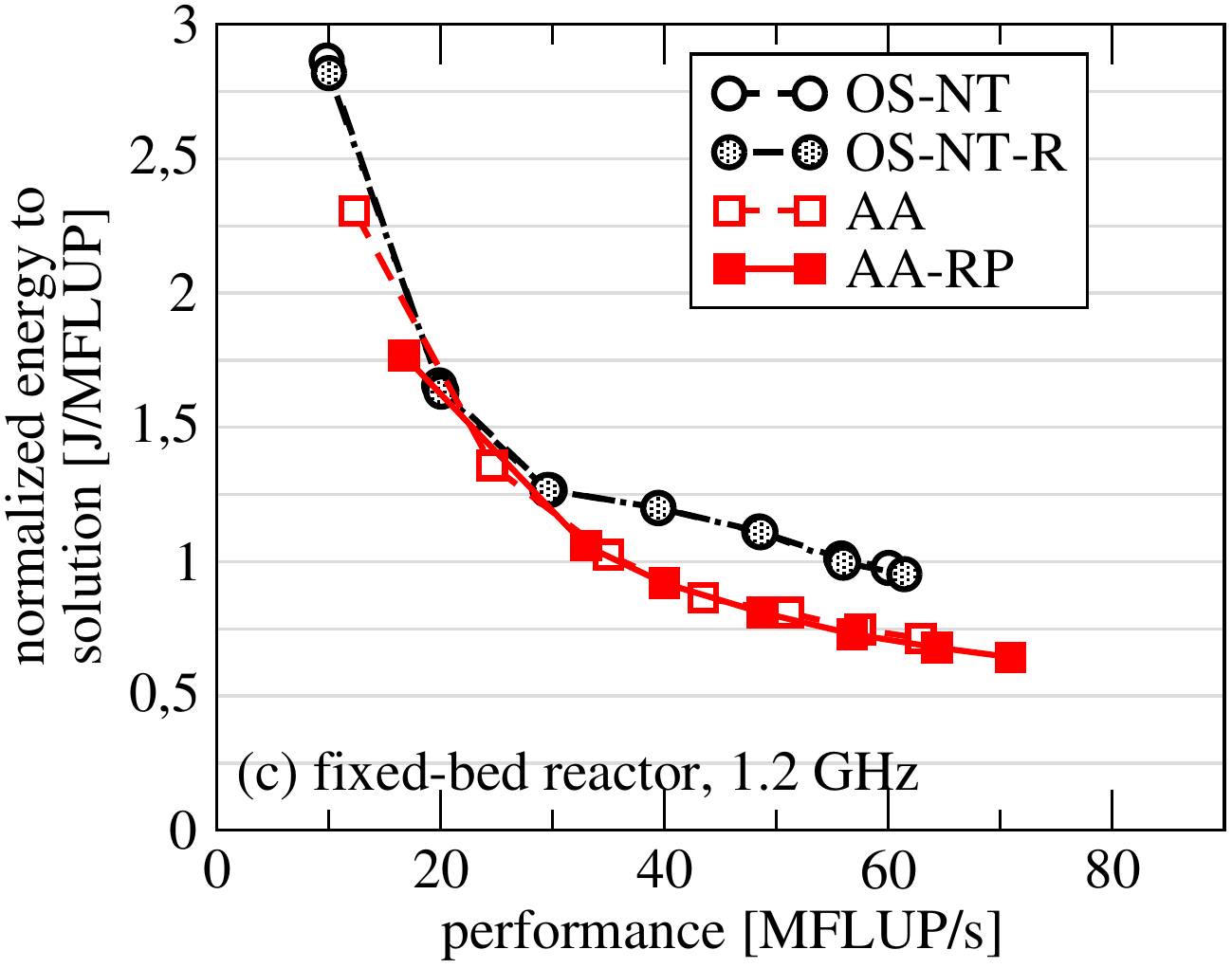}%
}
\subfloat{%
\label{fig:p:e4}%
\includegraphics[width=0.45\textwidth, clip=true]{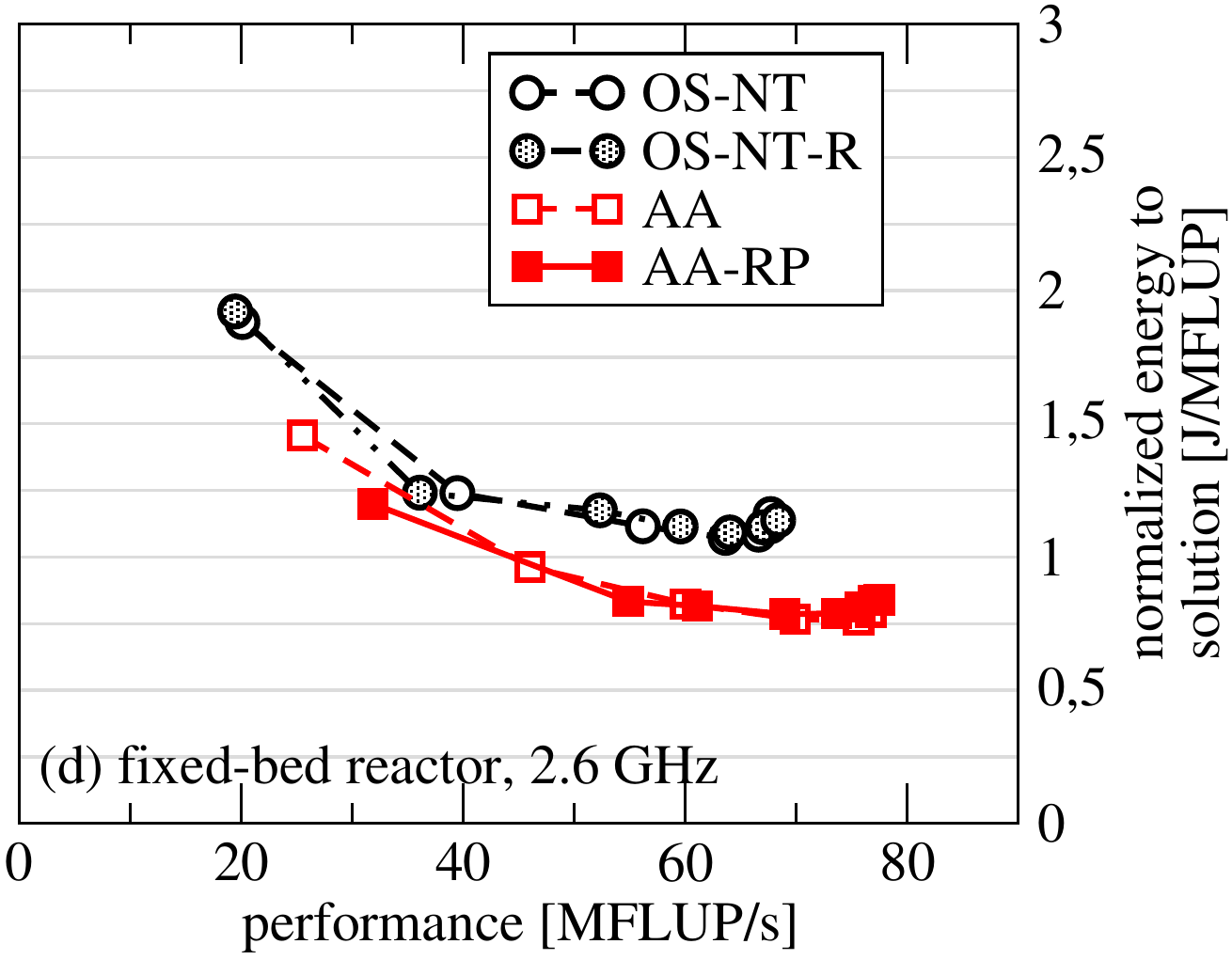}%
}%
\caption{Performance vs.\ Normalized energy to solution for OS-NT and AA in
the unoptimized and optimized variants, respectively.
The parameter on the curve is the number of cores used, ranging from one to
seven. 
}%
\label{fig:p:e}%
\end{figure}

For each benchmark run the power consumption of the core, uncore, and DRAM was
measured via likwid~\cite{likwid}.
Shown in Fig.~\ref{fig:p:e} is the normalized energy to solution (NETS), i.\,e.\
measured power divided by measured performance in the unit J/MFLUP.
The parameter on the curve is the number of cores utilized ranging from one to
seven.
OS-NT exhibits an interesting behaviour at $1.2$~GHz:
From three to five cores the NETS stays nearly constant while performance
increases.
With more cores NETS starts to decrease again.
This effect is unclear and deserves a more detailed investigation.
In all cases the NETS of AA is lower than of OS-NT, because of lower runtime
resulting from the higher performance.
The minimal NETS is reached, when performance saturates. 
If this point is reached when not all cores are utilized then each additional
core will only increase the NETS but not performance, visible in
Fig~\ref{fig:p:e} (b) for AA-RP.
The performance and NETS of AA-RP is always better than pure AA except for
the fixed-bed reactor at $2.6$~GHz.
Here at saturated performance nearly no difference is visible.
At most utilizing AA-RP in favor of AA reduces the NETS by around
$15$\.\%.

\subsection{Impact of Enumeration Function}
\label{sec:perf:enum}

\begin{figure}[tb]
  \centering
  \fbox{\includegraphics[width=0.75\textwidth, clip=true]{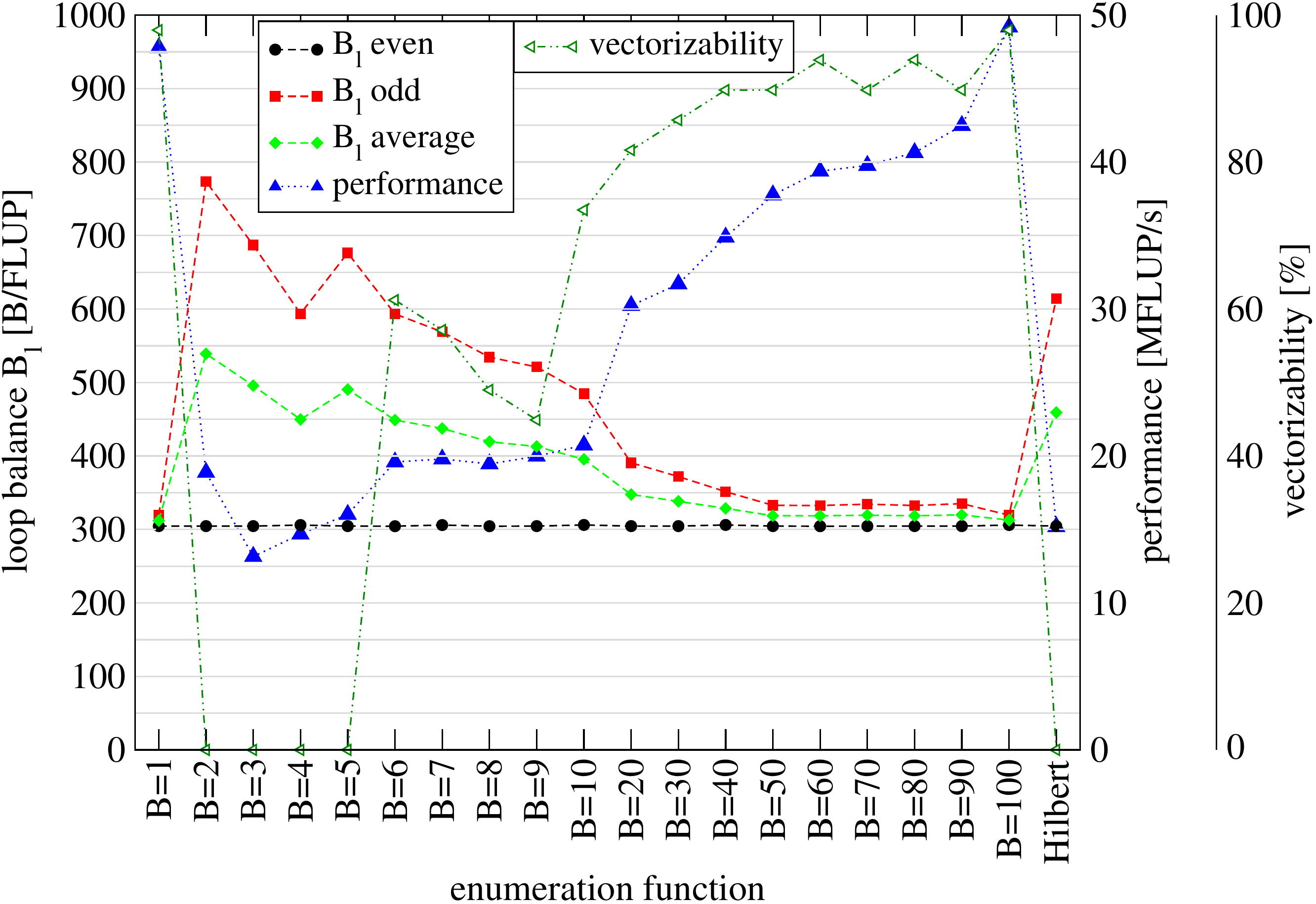}}
  \caption{Performance and in-cache loop balance $\xlb$ of the even and odd time
step of AA-RP on one core of the Haswell system.
As enumeration functions LS with different blocking factors $B$ as well as the
space filling Hilbert curve are utilized.  The loop balance was determined by
measuring the data traffic between the L1 and L2 cache with
likwid~\cite{likwid}.}
  \label{fig:enumfunctions}
\end{figure}

For the measurements conducted in the previous subsections LS
with blocking factor $B=1$, i.\,e.\ actual without blocking, was used.
Which enumeration function and parameters used for setting up the adjacency list
has direct influence on the resulting solver performance.
In Fig.~\ref{fig:enumfunctions} the loop balance $\xlb$ in cache and the
performance for \xaaa{} is plotted for LS with different blocking and the space
filling Hilbert curve.
Hereby the channel geometry and one core of the Haswell system ($2.6$~GHz) was
used.
The loop balance in cache was determined via measuring the data traffic between
L1 and L2 cache via likwid~\cite{likwid} for the even and odd time step
separately.
The in cache loop balance (black line) with $\xlb{} = 304$~B/FLUP of the even
time step is independent of the enumeration function, as here only accesses to
the local node are required, which can always be performed via direct
addressing.
Only the odd time step is affected (red line).
The plot shows that LS with and blocking factors ranging from $2 \le \xlb{} \le
10$ the performance (blue dashed line) drops to half the performance achieved
with $B=1$.
With this ordering cache lines during the odd time step are not efficiently
used.
They are evicted before all their containing PDFs have been used and therefore
must be reloaded.
The loop balance in this range nearly doubles for the odd time step.
This becomes the new bottleneck if too much time is spend with reloading cache
lines.
Furthermore the fraction of fluid nodes, which can be updated vectorized in the
odd time step, drops between $0$\,\% and $60$\,\%.
For larger blocking factors performance increases until with $B=100$ the $B=1$
performance is achieved again.
The blocking factors $B=1$ and $B=100$ can in this case be considered equally as
the diameter of the channel is $100$ nodes and they generate the same ordering.
Space filling curves like the Hilbert curve exhibit the same characteristics as
LS with small blocking factors: high in cache loop balance 
and low vectorizablility of the odd time step as well as low performance.
They suffer the same problem of inefficient cache line usage.

To reach a high single core performance LS with no blocking,
i.\,e.\ $B=1$, or with a high blocking factor is required.
Small blocking factors or space filling curves tend to inefficient cache line
usage and limit the vectorized update of fluid nodes which results in thereby a
lower performance.

\section{Large Scale Performance Aspects}
\label{sec:ls}

\begin{figure}[tb]
  \centering%
  \includegraphics[width=0.5\textwidth, clip=true]{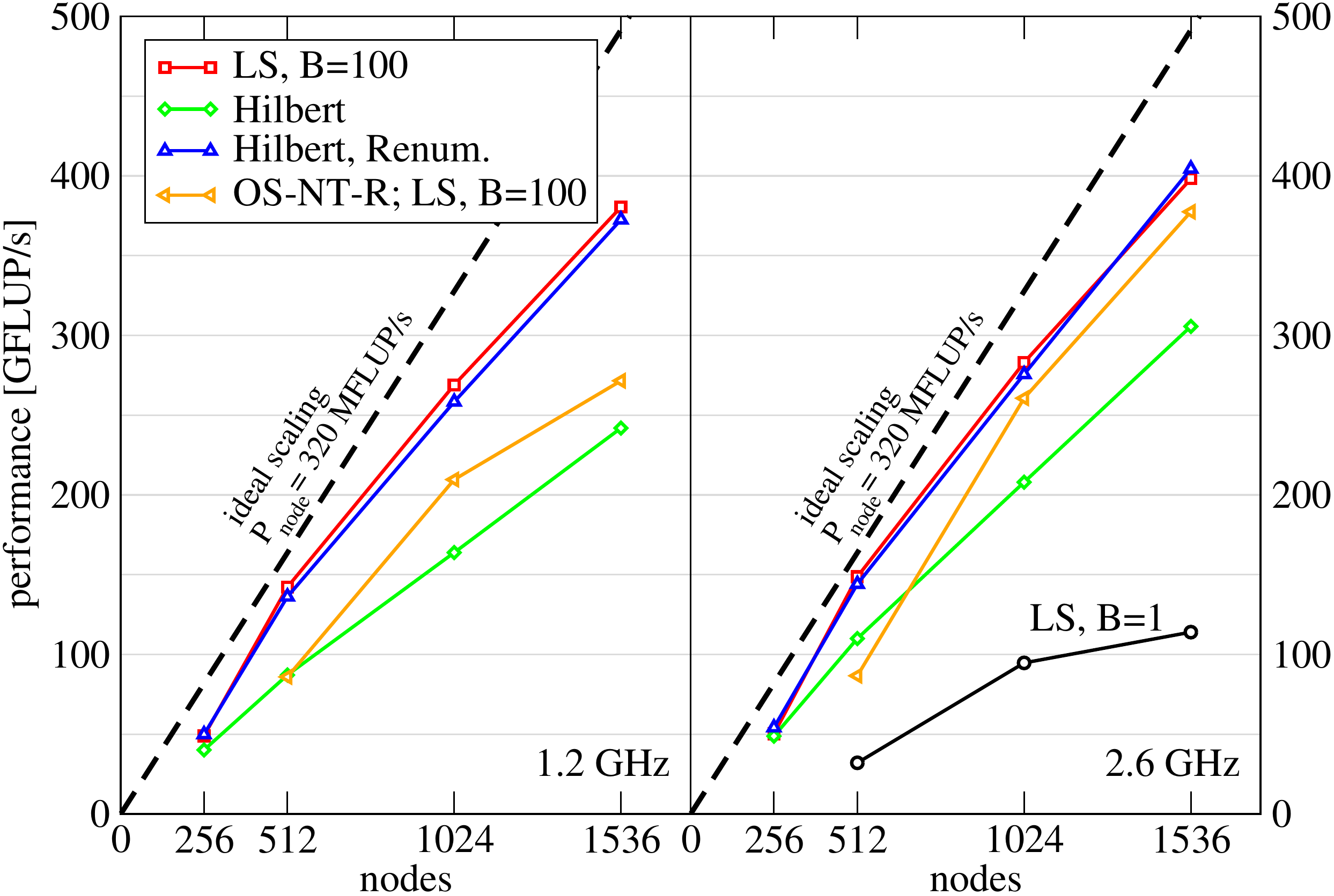}
  \caption{Performance of AA-RP with PPN$=28$ on the SuperMUC Phase~$2$ cluster.}
  \label{fig:ls:perf}
\end{figure}

For studying the strong scaling behaviour of \ilbdc{} a fixed bed reactor with
dimensions of $12500 \times 2500 \times 2500$ nodes is used.
It contains around $34 \times 10^9$ fluid nodes, requiring $4.8$~TiB for PDFs
and $2.3$~TiB for the adjacency list.

The performance results, when all physical cores of a compute node, i.\,e.\
PPN$=28$, are used are shown in Fig.~\ref{fig:ls:perf}.
When AA-RP and LS with blocking factor $B=1$ at $2.6$~GHz (right panel) is used
a very low performance is reached.
This is due to the high number of ghost PDFs, which must be exchanged between
neighboring partitions.
As with this blocking factor, the simulation domain is in principal cut
perpendicular to the x direction.
Each partition consists of a low number of slices which exhibit a high
communication volume.
At $256$ nodes around $300$~MiB per partition must be communicated, which drops
to around $60$~MiB for $1536$ nodes.
Through this high number of ghost cells per partition the memory of $256$ nodes
is exceeded, which is why there could be no measurement conducted.
By choosing a higher blocking factor, e.\,g.\ $B=100$, this can be mitigated.
Here the communication volume per partition decreases to around and the overall
performance increases.
For $256$ nodes only $8$~MiB per partition must be communicated.
This values decreases linearly to around $2$~MiB at $1536$ nodes.
Using a space filling curve enumeration functions like the Hilbert curve leads
also to partitions with a small surface.
The communication volume per partitions is nearly the same as with LS and
$B=100$.
But as already shown in Sect.~\ref{sec:perf:enum} they only achieve a poor single
core performance, which in turn leads to a low total performance.
For reference also the performance of OS-NT-R with LS and $B=100$ is plotted.
Hence two grids are required only at $512$ nodes enough memory was available.
As the communication properties are the same as with AA-RP and LS with $B=100$
the lower performance only results of the slower single core performance.

Reducing the clock speed to $1.2$~GHz (left panel of Fig.~\ref{fig:ls:perf})
only decreases the performance marginally when AA-RP and LS with $B=100$ is used.
Instead for AA-RP with Hilbert and OS-NT-R with LS and $B=100$ a decrease of
$20$\,\% and $30$\,\% is shown, respectively.
This stems to a large part from the lower single node performance.

At first sight super linear speedup is visible from the first to the second
point on each curve.
This is actually not the case, as the first points suffer from NUMA issues.
Allocation of the lattices and adjacency lists is not synchronized. 
If a process' local NUMA LD has not enough free memory, when the lattice and the
list is placed, because other processes have not already deallocated their
temporary structures needed for setup, parts of other LDs are used.
This decreases the single node performance as the bandwidth to a remote NUMA LD
is lower than to the local one.

A high performance for large scale simulations depends on the choice of the
right enumeration function and their parameter(s).
On the one hand a high blocking factor is required to achieve a good single core
performance, whereas on the other hand a small blocking factor (or an SFC)
generate partitions with small surfaces.
The goal is here to find a mechanism, which fulfills both conditions.

\subsection{Renumbering}

To achieve this goal a two step process is used.
In the first step an SFC, like the Hilbert curve, is initially used as
enumeration function.
This ensures, that during partitioning the communication volume of each
partition is minimized.
In the second step the fluid nodes assigned to a solver process are
\text{renumbered}.
Here LS with blocking factor $B=1$ can be chosen, for a high
single core performance.
The performance achieved is shown in Fig.~\ref{fig:ls:perf} as ``Hilbert
renum.''
With this method nearly the same performance as with LS and the
manually determined good parameter $B=100$ is achieved.

\subsection{Reducing the Number of Processes per Node}

\begin{figure}[tb]
\centering%
\includegraphics[width=0.45\textwidth, clip=true]{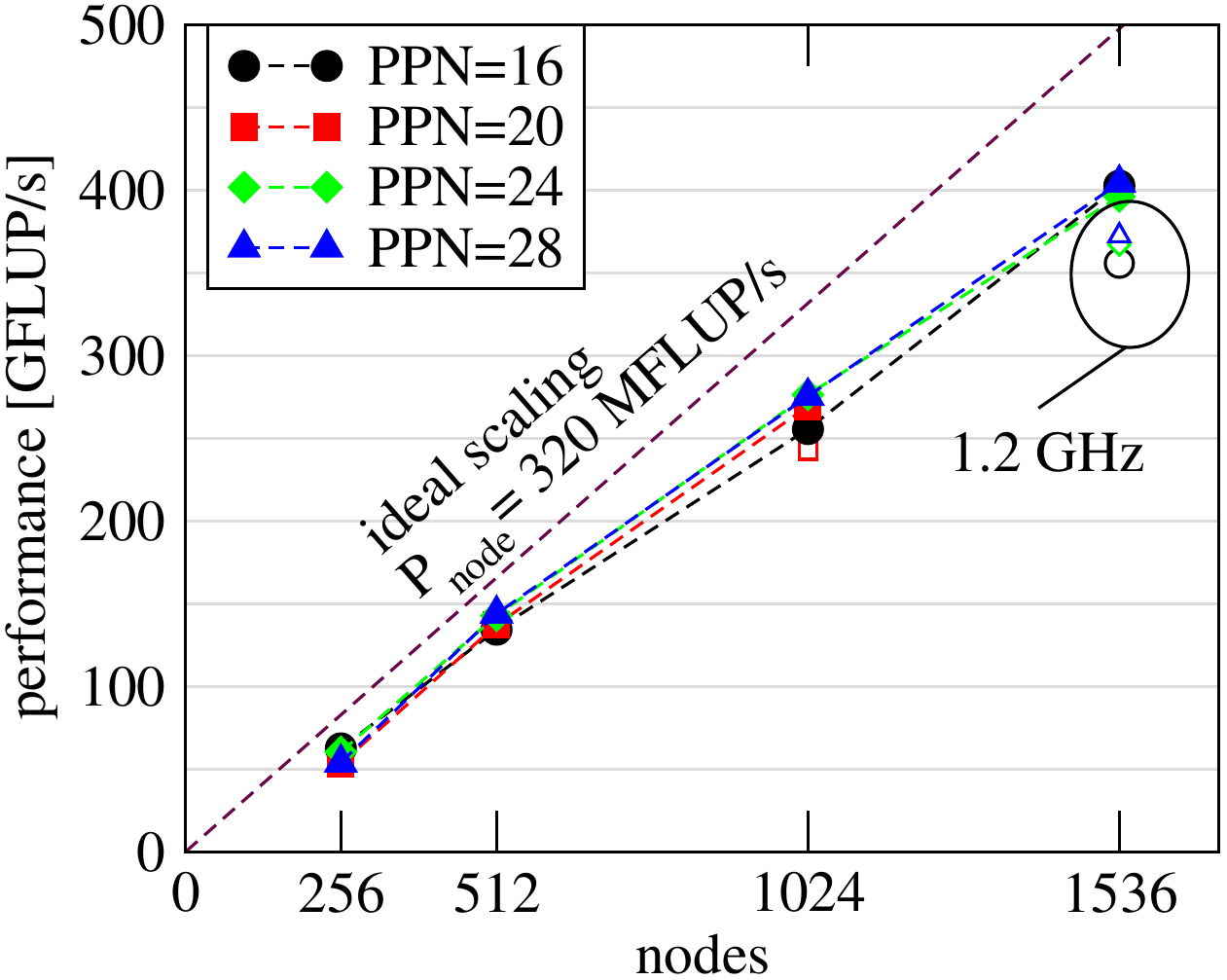}%
    \caption{Performance for AA-RB with Hilbert renumbered on the SuperMUC
Phase~$2$ cluster at $2.6$~GHz for different number of processes per nodes (PPN).
    }%
\label{fig:ls:ppn}%
\end{figure}

With AA-RP only three to four cores are required to saturate the performance, as
shown in Sect.~\ref{sec:perf:results}.
This is also the case in large scale.
In Fig.~\ref{fig:ls:ppn} the performance of AA-RP with Hilbert renumbered is
shown then the number of processes per node (PPN) is reduced from $28$ over $24$
and $20$ down to $16$.
As expected no performance impact is observed. 
Also with $1.2$~GHz this behaviour is the same.
Here the measurements from Fig.~\ref{fig:p:p} (c) would indicate otherwise, as
actually all cores are necessary to saturate performance.
This behaviour is unclear and requires a more detailed elaboration.

\subsection{Energy efficiency}

\begin{figure}[tb]
\centering%
\includegraphics[width=0.45\textwidth, clip=true]{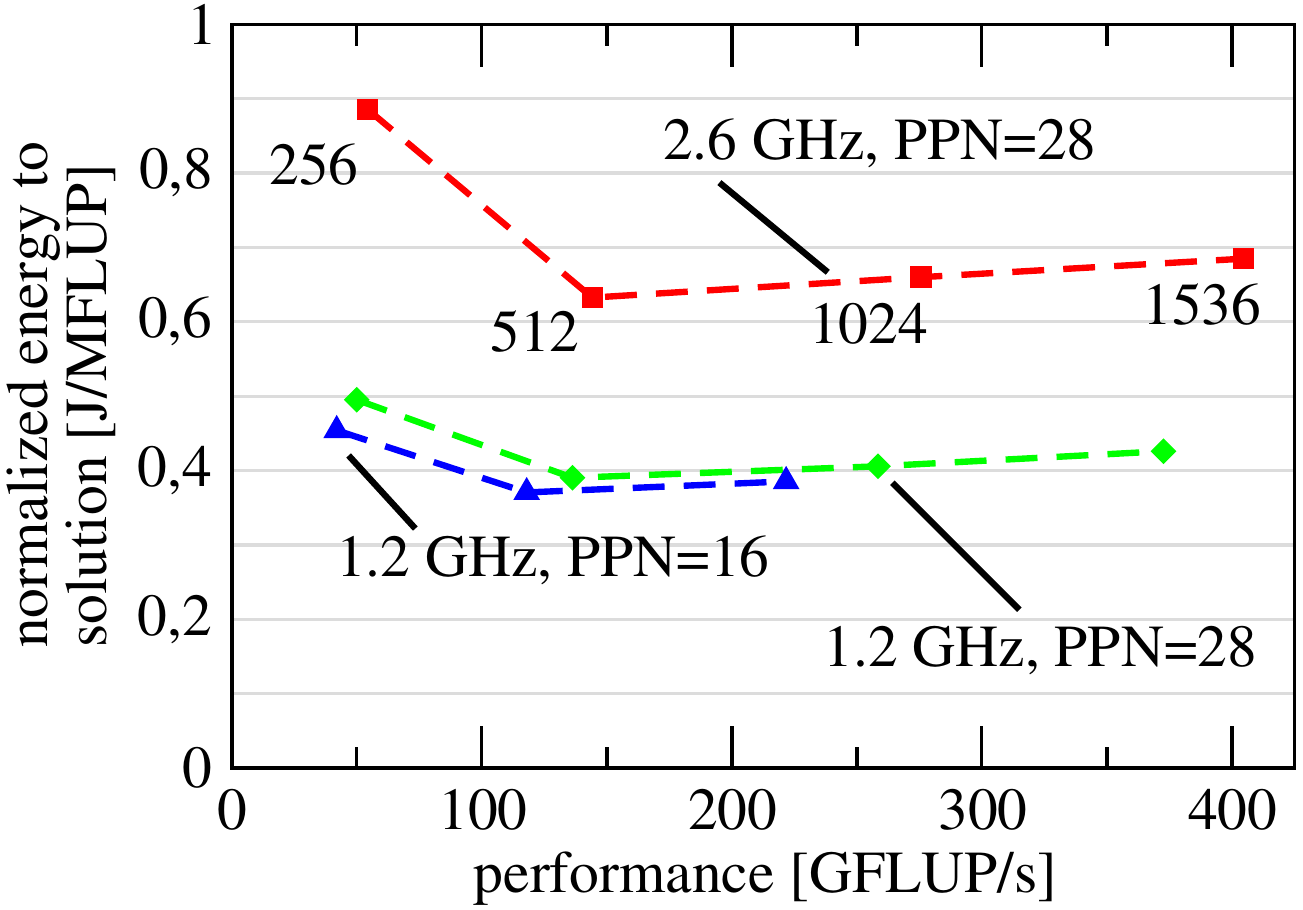}%
    \caption{Performance in relation to the normalized energy to solution (NETS)
for AA-RB with Hilbert renumbered on the SuperMUC Phase~$2$ cluster.
}%
    \label{fig:ls:energy}%
\end{figure}

At $1.2$~GHz AA-RP nearly reaches the performance of $2.6$~GHz also in the large
scale case, if the correct enumeration function is chosen.
Here energy savings of $40$\,\% can be achieved with the Hilbert renumbered
enumeration function, as shown in Fig.~\ref{fig:ls:energy}.
Reducing the number of processes on each node, i.\,e.\ reducing the number of
active cores, from PPN$=28$ down to PPN$=16$ only a small fraction of energy is
saved.
This is the same behaviour as in one NUMA LD for the fixed-bed reactor
at $1.2$~GHz as shown in Fig.~\ref{fig:p:e} (c).
Note that the value for $1536$ nodes in Fig.~\ref{fig:ls:energy} for $1.2$~GHz
and PPN$=16$ is missing.

\section{Conclusion}
\label{sec:conclusion}

Reduced indirect addressing (RIA) avoids for certain nodes the
indirect access and thereby reduces the data traffic per node update, i.\,e.\
the loop balance.
The implementation of RIA for the propagation step OS-NT showed that 
depending on the simulation geometry a performance increase of $15$\% is
observable.
For the AA-pattern with RIA no significant performance increase is visible, as
savings in the loop balance are less than for OS-NT with RIA.
AA-pattern with RIA enables the partial vectorization of the odd time step,
which has more impact on performance.
Hereby the loop balance is not altered and thus the saturated performance is not
higher than with only RIA enabled, but saturation can be reached with less
cores.
With lower frequencies like $1.2$~GHz instead of $2.6$~GHz no saturation
occurs.
At this point partial vectorization increases the overall performance by up to
$30$\% depending on the simulation geometry.
The savings in the energy consumption (up to $30$\,\%) are higher with simple
structured simulation geometries as there the benefit from partial vectorization
is higher.

In the large scale case choosing the right enumeration function is crucial.
With a simple two step method we avoid the manual search for good parameters,
which depend on the simulation geometry, and still reach the same performance.
Also here are the single core optimizations visible, which allow to run at lower
frequencies and use less cores.
Here up to $40$\,\% of energy savings are possible.

We would like to thank the Leibniz Computing Center (LRZ) in Garching, Germany
for the great support during conducting the benchmarks.
This work was financially supported by KONWIHR III and partially supported by
BMBF under grant No.\ 01IH08003A (project SKALB).

\bibliographystyle{abbrv}
\bibliography{ISC-2016}

\end{document}